\documentclass[a4paper,12pt]{article}
\usepackage{comment}

\usepackage{amssymb}

\usepackage{natbib}
\bibpunct
{(} {)} {,} {a} {} {;} 

\usepackage{graphicx, ifthen}
\usepackage[hidelinks]{hyperref}

\newcommand{\citetbjps}[2][]{\ifthenelse{\equal{#1}{}}{\citeauthor{#2} ([\citeyear{#2}])}{\citeauthor{#2} ([\citeyear{#2}], #1)}}
\newcommand{\citealtbjps}[2][]{\ifthenelse{\equal{#1}{}}{\citeauthor{#2} [\citeyear{#2}]}{\citeauthor{#2} [\citeyear{#2}], #1}}
\newcommand{\citepbjps}[2][]{\ifthenelse{\equal{#1}{}}{(\citeauthor{#2} [\citeyear{#2}])}{(\citeauthor{#2} [\citeyear{#2}], #1)}}
\newcommand{\citeyearbjps}[2][]{\ifthenelse{\equal{#1}{}}{[\citeyear{#2}]}{[\citeyear{#2}], #1}}
\newcommand{\citeyearparbjps}[2][]{\ifthenelse{\equal{#1}{}}{([\citeyear{#2}])}{([\citeyear{#2}], #1)}}
\newcommand{\citeposbjps}[2][]{\ifthenelse{\equal{#1}{}}{\citeauthor{#2}'s ([\citeyear{#2}])}{\citeauthor{#2}'s ([\citeyear{#2}], #1)}}

\usepackage{newtxtext,newtxmath}
\usepackage{microtype}

\usepackage{sectsty} \allsectionsfont{\large \centering} \makeatletter
   \def\@seccntformat#1{\csname the#1\endcsname.\quad}
\makeatother

\usepackage{tocloft}

\cftpagenumbersoff{section}

\begin{document}

\title{\textbf{On the Mathematics and Metaphysics of the Hole Argument}}

\author{\textbf{Oliver Pooley
and James Read}}
\date{}

\maketitle

\begin{abstract}
We make some remarks on the mathematics and metaphysics of the hole argument, in response to a recent article in this journal by \citetbjps{Weatherall}.
Broadly speaking, we defend the mainstream philosophical literature from the claim that correct usage of the mathematics of general relativity `blocks' the argument.
\end{abstract}

\vspace{-1cm}
{\em \tableofcontents}

\section{Introduction}

Consider models of general relativity $\mathcal{M} = \left( M, g_{ab} \right) $
and $\widetilde{\mathcal{M}} = \left( M, \tilde{g}_{ab} \right) $,\footnote{In order to guard against conflating such models with the possibilities they represent (a distinction that is vital when discussing the hole argument), we do not refer to the models as `spacetimes'.} both foliable into spacelike hypersurfaces, identical up to some such hypersurface $\Sigma$, but differing by a diffeomorphism (such that $\tilde{g}_{ab} = d_* g_{ab}$, where $d_*$ represents the push-forward map) which acts non-trivially to the future of $\Sigma$. Na\"{i}vely, if $\mathcal{M}$ is interpreted as representing some possible world \(W\), then $\widetilde{\mathcal{M}}$ is to be interpreted as representing a distinct possible world $\widetilde{W}$, in which (the images of---that is,~the physical correlates of) both the metric field $g_{ab}$ and material fields are distributed differently (but isomorphically) on (the image of) $M$. The following problem now arises: given (the history associated with) these solutions of general relativity up to $\Sigma$, the theory (it seems) simply does not determine which of $W$ or $\widetilde{W}$ will be realized. Thus, the theory appears to be radically indeterministic.

The foregoing is a version of the hole argument, first presented by Einstein in 1913 as a means of excluding from consideration any diffeomorphism invariant theory of gravitation\footnote{We understand the notion of `diffeomorphism invariance' to be distinct from that of `general covariance'---see \citealtbjps{Pooley2017}.}---an argument that he would later reject. The argument was revived by \citetbjps{EN} as a means of rejecting substantivalism about the manifold $M$ of general relativity---for, as they elaborate, the above problem can be avoided if the independent reality of (the physical correlate of) $M$ is rejected, so that one can no longer articulate the difference between $W$ and $\widetilde{W}$, and so cannot articulate the distinction that gives rise to the apparent indeterminism. Earman and Norton's paper spawned a sizeable literature, including several defences of the viability of substantivalism.\footnote{For more on contemporary philosophy of physics discussion of the hole argument, see \citealtbjps{Norton2011}, \citealtbjps{Pooley2013}, \citealtbjps{PooleyKW}, \citealtbjps{RW} and references therein.}

In a recent paper, Weatherall has questioned whether this second, philosophical wave of writing on the hole argument is well-motivated. His central thesis is articulated at the beginning of his article:
\begin{quote}
    Einstein and the generations of physicists and mathematicians after him were right to reject the hole argument. It is based on a misleading use of the mathematical formalism of general relativity. If one is attentive to mathematical practice, I will argue, the hole argument is blocked. \citepbjps[p.~330]{Weatherall}
\end{quote}
Later, Weatherall extols an apparent virtue of his approach:~`This  particular option is distinctive---and, I think, attractive---because it is essentially neutral on the metaphysics of space and time.' \citepbjps[p.~330]{Weatherall}

Our purpose in this article is to make some constructive remarks on the role of mathematics and metaphysics in the context of the hole argument. Broadly speaking, these lead us to reject Weatherall's approach: a proper understanding of the mathematics of general relativity does not suffice to block the hole argument.

\section{Indeterminism versus Underdetermination}

In the guise presented above, the hole argument raises the spectre of indeterminism:~given initial data on and prior to $\Sigma$, general relativity is silent on whether the world will evolve as per $\mathcal{M}$, or as per $\widetilde{\mathcal{M}}$. This is, indeed, the version of the argument most often discussed in the literature. One may also, however, use the hole construction to generate (what we call for the purposes of this paper) a problem of underdetermination.\footnote{The distinction between the indeterminism and underdetermination versions of the hole arguments is closely related to Earman and Norton's \citeyearparbjps{EN} distinction between the `indeterminism dilemma' and the `verificationist dilemma'. Our thanks to John Norton and Neil Dewar for discussion of how best to understand the underdetermination version of the argument.}

Suppose that model $\mathcal{M}$ is a candidate to represent the actual world. Now consider the model $\widetilde{\mathcal{M}}$, generated from $\mathcal{M}$ via a hole  diffeomorphism $d$. Since all relational quantities (Einstein's `point coincidences'---cf.~\citepbjps[\S8.2]{Norton2011}) are preserved under such a diffeomorphism, it appears that the empirical data available to any observer within a world represented by either $\mathcal{M}$ or $\widetilde{\mathcal{M}}$ will fail to determine which model represents that observer's world.

In particular, suppose that, according to $\mathcal{M}$, the observer at the salient stage of their trajectory is located at (the spacetime point represented by) $p$ and that $d$ maps $p$ to a distinct point $q$. According to $\widetilde{\mathcal{M}}$, therefore, the relevant stage of the observer's trajectory is located at $q$. It follows that no measurement that the observer might perform at that point along their trajectory can determine whether they are located at (the point represented by) $p$ or at (the point represented by) $q$, for the outcomes of any measurements are the same according to $\mathcal{M}$ and $\widetilde{\mathcal{M}}$.

When Weatherall presents the hole argument at \citepbjps[\S3]{Weatherall}, he does not distinguish clearly this underdetermination version of the argument from the indeterminism version. As we show in what follows, some of Weatherall's criticisms of the hole argument do not survive proper attention to this distinction.

\section{Weatherall's Arguments}

Weatherall can be read as providing two distinct lines of criticism of the hole argument, although they are not individuated explicitly as such in his paper. (We recognize that by regimenting Weatherall's criticism in this way, we may thereby fail to represent it completely faithfully; nevertheless, we believe that this is a productive way to proceed.)

Central to both arguments is a distinction that Weatherall draws between $1_M$ and $\tilde{\psi}$, two maps between $\mathcal{M}$ and $\widetilde{\mathcal{M}}$, which he introduces as follows:
\begin{quote}
    Assertions concerning point $p$ in the context of both $\left( M, g_{ab} \right)$ and $\left( M, \tilde{g}_{ab} \right)$ are implicitly made relative to the identity map on this manifold, $1_M \colon M \rightarrow M$. Meanwhile, we have the map $\psi \colon M \rightarrow M$, which is a diffeomorphism. It is $\psi$ that gives rise to the isometry 
    $\tilde{\psi} \colon \left( M ,g_{ab}\right) \rightarrow \left( M , \tilde{g}_{ab} \right)$, where I have again used a [tilde] to  be clear that $\psi$ and $\tilde{\psi}$ are different maps---$\psi$ is an automorphism of $M$, whereas $\tilde{\psi}$ is an isomorphism, but not an automorphism, between the Lorentizan manifolds $\left( M, g_{ab}\right)$ and $\left( M, \tilde{g}_{ab}\right)$. \citepbjps[p.~336]{Weatherall}
\end{quote}
After introducing the distinction,  Weatherall goes on to consider the comparisons of two hole-diffeomorphic models under each of the maps:
\begin{quote}
    When we say that $\left( M, g_{ab}\right)$ and $\left( M, \tilde{g}_{ab}\right)$ are isometric spacetimes, and thus that they have all of the same invariant, observable structure, we are comparing them relative to $\tilde{\psi}$. Indeed, we must be because~[\ldots \!] there is no sense in which $1_M$ either is or gives rise to an isometry. In other words, relative to $1_M$, $\left( M, g_{ab}\right)$ and $\left( M, \tilde{g}_{ab}\right)$ are not equivalent, physically or otherwise. The reason is that there exist points $p \in O$ [the region on which the diffeomorphism $d$ acts non-trivially] at which ${\left( g_{ab} \right)}_{|p} \neq {\left( \tilde{g}_{ab} \right)}_{|1_M \left(p\right)}$.  Consider an observer sitting at a point of spacetime represented by the point $p \in O$ of $\left( M, g_{ab}\right)$. If one were to attempt to represent that same observer's location by the point $1_M\left( p\right) = p$ of $\left( M, \tilde{g}_{ab}\right)$, one would conclude that the observer would see measurably different metrical properties, different curvature, and so on. In general, only one of these assignments can be correct. Meanwhile, if one only considers $\tilde{\psi}$, no disagreement arises regarding the value of the metric at any given point, since for any point $p\in M$, ${\left( g_{ab} \right)}_{|p} = {\left( \tilde{g}_{ab} \right)}_{|\psi\left( p \right)}$ by construction. \citepbjps[p.~336]{Weatherall}
\end{quote}

Weatherall's reasoning in this passage can be summarized as follows:\ when $\mathcal{M}$ and $\widetilde{\mathcal{M}}$ are compared using $1_M$ (that is, when one's interpretation of these models is sensitive to differences captured by $1_M$ and so to questions concerning the field values at any given $p \in M$), they should be taken to represent observationally distinguishable situations. On the other hand, when $\mathcal{M}$ and $\widetilde{\mathcal{M}}$ are compared using $\tilde{\psi}$ (that is, when one's interpretation of these models is sensitive only to differences captured by $\tilde{\psi}$ and so \emph{not} sensitive to questions concerning the field values at any given $p \in M$), then the models are to be taken to be physically and not just empirically equivalent.

We will call Weatherall's two arguments against the hole argument `the equivocation argument' and `the argument from mathematical structuralism'. According to the equivocation argument, while comparisons of models either via $1_M$ or via $\tilde{\psi}$ might be legitimate, the hole argument rests on an illegitimate equivocation between the two. According to the argument from mathematical structuralism, models should be compared via $\tilde{\psi}$, but the hole argument requires an (illegitimate) comparison of models via $1_M$.

We are not persuaded by either argument. For convenience, we summarize here the three central threads of our critique, which are covered in detail in the following sections. The first two strands focus on the equivocation argument. The third addresses the argument from mathematical structuralism.

\begin{enumerate}
    \item Weatherall's claim that the hole argument involves an equivocation leaves the indeterminism version of the argument completely untouched. This is because $1_M$ alone is used to articulate this version of the argument. Comparisons via $\tilde{\psi}$, and the related claim that models so compared are observationally equivalent, play no role in the indeterminism version of the hole argument. The charge of equivocation is therefore at best relevant to the underdetermination version of the hole argument.
    \item Weatherall's suggestion that hole-diffeomorphic models are observationally inequivalent when compared using $1_M$ is questionable. His position is either a close cousin of Maudlin's response to the static shift argument against Newtonian absolute space or it presupposes a controversial conception of empirical equivalence. Either way, Weatherall's alleged dissolution of the underdetermination problem is contentious.
        \item Weatherall's claim that the hole argument is blocked when models are compared using $\tilde{\psi}$ is also questionable, for the interpretation of models that such a comparison might mandate does nothing, by itself, to eliminate the relevant space of metaphysical possibilities.\end{enumerate}

\section{The Equivocation Argument}

According to Weatherall, in order to generate the hole argument, one has to compare $\mathcal{M}$ and $\widetilde{\mathcal{M}}$ using $1_M$, insofar as one regards them as representing physically distinct states of affairs, but {also}---in an (allegedly) illegitimate manner---compare them using $\tilde{\psi}$, insofar as one wishes to regard the worlds represented by those models as observationally indistinguishable. He thus maintains that the argument rests 
on an equivocation:
\begin{quote}
    There is a sense in which $\left( M, g_{ab} \right)$ and $\left( M, \tilde{g}_{ab} \right)$ are the same, and there is a sense in which they are different [\ldots] But---and this is the central point---one cannot have it both ways. Insofar as one wants to claim that these Lorentzian manifolds are physically equivalent, or agree on all observable/physical structure, one has to use $\tilde{\psi}$ to establish a standard of comparison between points. And relative to this standard, the two Lorentzian manifolds agree on the metric at every point---there is no ambiguity, and no indeterminism. (This is just what it means to say that they are isometric.) Meanwhile, insofar as one wants to claim that these Lorentzian manifolds assign different values of the metric to each point, one must use a different standard of comparison. And relative to this standard---that given by $1_M$---the two Lorentzian manifolds are not equivalent. One way or the other, the hole argument seems to be blocked. \citepbjps[pp.~338--39]{Weatherall}
\end{quote}

Let us assess this equivocation argument. Recall Weatherall's central contentions concerning comparisons of models via $1_M$:
\begin{quote}
[R]elative to $1_M$, $\left( M, g_{ab}\right)$ and $\left( M, \tilde{g}_{ab}\right)$ are not equivalent, physically or otherwise. The reason is that there exist points $p \in O$ at which ${\left( g_{ab} \right)}_{|p} \neq {\left( \tilde{g}_{ab} \right)}_{|1_M \left(p\right)}$.
\end{quote}
So far, this observation is as much an elaboration of the hole argument as a response to it. Regarding $\mathcal{M}$ and $\widetilde{\mathcal{M}}$ as physically inequivalent (that is, as representing distinct possibilities) when compared via $1_M$ is central to both arguments. On the indeterminism argument, the claim is that the laws together with the state of spacetime outside of $O$ (that is, outside of the `hole') fail to fix whether the state of $O$ is as described by $\mathcal{M}$ or by $\widetilde{\mathcal{M}}$. On the underdetermination argument, the claim is that no observation could tell us whether the state of $O$ is as described by $\mathcal{M}$ or by $\widetilde{\mathcal{M}}$. That ${\left( g_{ab} \right)}_{|p} \neq {\left( \tilde{g}_{ab} \right)}_{|1_M \left(p\right)}$ is precisely the point of both arguments: amongst the things that the laws do not fix, or that observations cannot disclose, are, \emph{inter alia}, which particular geometrical properties get to be instantiated by (the spacetime point represented by) $p$.

Weatherall, however, goes on:
\begin{quote}
Consider an observer sitting at a point of spacetime represented by the point $p \in O$ of $\left( M, g_{ab}\right)$. If one were to attempt to represent that same observer's location by the point $1_M \!\left( p\right) = p$ of $\left( M, \tilde{g}_{ab}\right)$, one would conclude that the observer would see measurably different metrical properties, different curvature, and so on. In general, only one of these assignments can be correct.
\end{quote}
At this point Weatherall's failure to distinguish clearly between the indeterminism and underdetermination versions of the hole argument becomes significant. His claim that at most one of $\mathcal{M}$ and $\widetilde{\mathcal{M}}$ correctly characterizes the observations made by an observer inside the hole, far from blocking the indeterminism version of the argument, in fact makes it more acute! For, if Weatherall is to be believed, what the matching regions of the two models and laws fail to fix are not merely (i) metaphysical but physically undetectable facts concerning which spacetime points have which properties (as many have thought), but rather (ii) states of affairs that are observationally distinguishable.

In light of Weatherall's apparent failure to spot this problem, a few features of the indeterminism version of the hole argument are worth re-emphasizing. Precisely because the diffeomorphism $d$ acts trivially outside of $O$, it is uncontroversial that spacetimes represented by $\mathcal{M}$ and $\widetilde{\mathcal{M}}$ are to be regarded as completely---not just empirically---identical in the region outside of $O$.\footnote{This is not to deny that one could concoct contexts where $\mathcal{M}$ and $\widetilde{\mathcal{M}}$ are deployed to represent spacetimes that differ on $M \setminus O$. The hole argument, however, is clearly not such a context.} Regarding $\mathcal{M}$ and $\widetilde{\mathcal{M}}$ as equivalent outside of $O$, but as differing within $O$, therefore does not involve any allegedly illegitimate equivocation between $1_M$ and $\tilde{\psi}$, for $1_M$ and $\tilde{\psi}$ agree completely outside of $O$; only comparisons via $1_M$ ever need be considered. Weatherall, moreover, appears to concede, at least for the sake of argument, that comparisons via $1_M$ are legitimate in themselves and agrees that, so compared, the models are not physically equivalent, because of their different attributions to the points in $O$. He therefore seems to walk headlong into the indeterminism version of the hole argument.

To the extent that comparisons via $\tilde{\psi}$ are sometimes invoked in the context of the indeterminism argument, it is after the event, in order to downplay the apparent problem: perhaps the theory is, strictly speaking, indeterministic but the facts not fixed are esoteric metaphysical matters concerning point identities; they are not facts that affect the theory's observable predictions (for consider $\tilde{\psi}$!). Weatherall, however, appears to view $\mathcal{M}$ and $\widetilde{\mathcal{M}}$ as not even empirically equivalent if compared via $1_M$. It is this claim, and the underdetermination version of the hole argument, to which we now turn.

We start by noting one sense in which Weatherall's critique fails to engage directly with what we take to be the natural way to pose the underdetermination problem. When introducing the problem above, we considered (like Weatherall) an observer located within the region $O$ where the hole diffeomorphism acts non-trivially. Idealizing somewhat, suppose once again that, {as it occurs in model $\mathcal{M}$}, $p \in M$ represents the observer's location at the moment of the relevant observation event. Our contention is that in model $\widetilde{\mathcal{M}}$ it is then $d(p) = q \neq p$ that represents the same observer's location at the same moment of their trajectory. To assert this is {not} to switch illegitimately from comparing the models via $1_M$ to comparing them via $\tilde{\psi}$. It is simply to recognize that the relevant fields representing the observer's trajectory, their measuring instruments and the results of the measurements have been dragged by $d$ from $p$ in $\mathcal{M}$ to $q$ in $\widetilde{\mathcal{M}}$. $1_M$ remains the official map by which we identify {points} across models. That does not by itself imply, however, that it can also be used to identify the perspectives of observers across models.

As Weatherall notes in a footnote to the passage quoted \citepbjps[p.~336, fn.~20]{Weatherall}, his suggestion that, when compared via $1_M$, $\mathcal{M}$ and $\widetilde{\mathcal{M}}$ should count as representing observationally discernible states of affairs echoes claims made by Tim Maudlin, in the context of the static shift argument against absolute space. (Recall that static shifts of a Newtonian world are time-independent rigid translations of the entire material content of the universe.) It is worth briefly reviewing Maudlin's position, for it involves two distinct claims. The first, though controversial, is one that we are inclined to accept; the second we reject. An analogue of the first claim can be used to block the underdetermination argument in the context of general relativity; the second claim, however, is no more defensible in the context of general relativity than it is in the context of Newtonian physics.

What, then, is Maudlin's position regarding static shifts? The claim with which we are sympathetic is that, even if one acknowledges that possible Newtonian worlds can differ merely by static shifts, the existence of this plurality of possibilities does not pose an epistemological challenge.\footnote{In contrast, as Maudlin emphasizes, the case of worlds differing by kinematic shifts, that is, differing by a time-independent and uniform change in the absolute velocities of all material bodies, does pose an epistemological challenge. Prior to Maudlin, variants of the same point were made by \citetbjps{Horwich}, \citetbjps{Field}, and \citetbjps{Teller}.} To paraphrase Maudlin, my acknowledging that there is a genuine distinction between a world in which my desk\footnote{That is, the desk belonging to the coauthor currently typing.} is here and a world in which it is three metres north of here (but in which all spatial relations between bodies remain just the same) does not mean that there are any positional facts of which I am ignorant. I cannot contemplate these two possibilities without knowing full well that the possibility corresponding to the actual world is the first. To make a closely related point in terms of a class of statically-shifted models apt to represent these possibilities: given an {arbitrary choice} of one of the models from this class to represent the actual world, I {know} that the worlds represented by the other models are not my own. (We stress that the claim that static shifts pose no epistemological problem has not gone unchallenged---see \citepbjps{Dasgupta2} for recent discussion.)

A sympathetic reading of Weatherall interprets him as making an analogous point in the context of the underdetermination argument. Suppose that we stipulate that $p$ is to be the point of $M$ that represents our observer at the relevant observation event. Then, relative to that stipulation, it will be determined that it is $\mathcal{M}$ (say) and not $\widetilde{\mathcal{M}}$ that correctly assigns to $p$ the geometrical properties that the observer detects. To put the point another way, there is, on reflection, something suspect about a claim that we made when originally framing the underdetermination problem, namely, that no measurement that an observer might perform at the relevant moment could determine whether they are located at (the point represented by) $p$ or at (the point represented by) $q$. For what determines which spacetime points are represented by $p$ and $q$? Plausibly, the only way that this reference relation can be tied down is for the observer (or us) to choose arbitrarily one such model as representing their world. Relative that choice, the other models will represent possibilities that the observer knows to be (for them) counterfactual.

As noted, we are sympathetic to this response to the underdetermination argument.\footnote{It is worth stressing that this Maudlin-style move is ineffective against the hole argument in its indeterminism form. Even if one can know by stipulation which model represents my actual future, it remains the case that the past and the laws do not determine that the future is as represented in this model rather than as represented, counterfactually, in hole-diffeomorphic models.} On our understanding of this response, although one is not ignorant of which possibility represented by the models is one's own, the possibilities are not {observationally} distinguishable. Maudlin, however, does claim that they are and, as we have seen, Weatherall follows his lead. We think that this is a mistake.

In arguing his case, Maudlin switches from the example of static spatial shifts to time translations. In order to catalogue efficiently our points of disagreement, we quote him at length:
\begin{quote}
A universe created 15 billion years ago is observationally distinguishable from one just like it (that is, having a qualitatively identical total history) which began within the last four minutes. Things would look awfully different if the big bang had occurred in the last half hour. Of course, if the big bang had occurred four minutes ago then in another 15 billion years there might be someone who looks just like me writing a sentence that looks just like this. But that person would have no difficulty determining that he is not alive now, just as I have no difficulty knowing that I will not be alive then. \citepbjps[p.~190]{Maudlin}
\end{quote}
We respond: would things look awfully different if the big bang had occurred in the last half hour? To whom? Things would look different to someone located, in this counterfactual universe, at the very time we find ourselves at right now, but there is no one in this world located at that time (since the world is hypothesized to have a qualitatively identical total history to that of the actual world and no observer was around four minutes after the big bang).

Maudlin concedes that, if the big bang had occurred four minutes ago then, in another 15 billion years, there might be someone who looks just like him writing a sentence just like the one quoted. But first note that, since this universe is qualitatively identical, it is not merely the case that there {might} be such a person---in fact, of course, there {will} be such a person. Secondly, and crucially, this person who looks just like Maudlin is none other than Maudlin himself, occupying a moment in time in this counterfactual possibility different to his actual temporal location.\footnote{At least, this is what haecceitists like Maudlin should say. Even a counterpart theorist will agree that the Maudlin-like individual of this counterfactual possibility underwrites \emph{de re} truths about what Maudlin himself would have observed had the universe been created only four minutes ago. In other words, the counterpart theorist agrees that, had things been as envisaged in the counterfactual scenario, Maudlin's experiences would have been qualitatively identical to his actual experiences.} So in this counterfactual world his observations, moment by moment, throughout his illustrious career, are qualitatively identical to his actual observations. The same goes for every other observer. {In this sense}, the worlds are observationally indistinguishable.  Maudlin may know that the world described is not the actual world, but this is not empirical knowledge; he did not come by this knowledge by checking experimentally in 1993 that the big bang had not happened only four minutes previously.

The parallel between Maudlin's position and Weatherall's is clear. Just as Maudlin contrasts the goings on at a fixed time to argue that his two worlds are observationally distinguishable, so Weatherall contrasts the field values at a fixed manifold point in the two models to argue that at most one of the worlds they represent correctly captures what an observer at that point would measure. There is also, however, an important difference between Maudlin's and Weatherall's positions that is worth dwelling on.

Maudlin's worlds---the actual world and its rigidly shifted counterparts---include all relevant observers. They do so because they are complete universes. Similarly, any mathematical models apt to represent such worlds will model all observers, and not just the content of their observations. In general, explicit representation, at least in an idealized manner, of the observer, as a {physical} system within spacetime, naturally leads to what one might call an `immanent' conception of empirical (in)equivalence:
\begin{quote}
Two models are empirically distinct just in case there are relevant relational differences between the field configurations in each.\footnote{Of course, identifying the `relevant relational differences' which contribute to such empirical differences is a non-trivial matter. See for example \citepbjps{TMNJR} for discussion.}
\end{quote}
In our view, this is the notion of empirical equivalence most relevant to discussions of the hole argument. It is the notion of empirical equivalence presupposed in \citepbjps[pp.~521--22]{EN}. It is in accord with Stein's call to `schematize the observer' \citeyearparbjps{Stein}.\footnote{As Curiel writes (in the context of an extended discussion of Stein), `At bottom, then, what secure epistemic content a scientific theory has must rest in large part on the meanings expressed in the sound articulation of experimental knowledge, for that is the final arbiter of empirical success. This requires at a minimum that we be able, at least in principle, to construct appropriate and adequate representations of actual experiments and observations in the frameworks of our best scientific theories, that is, representations of physical systems and experimental apparatus in relation to each other as required by actual experiments, not just representations of physical systems \emph{simpliciter}, in abstraction from experimental practice' \citepbjps[p.~10]{Curiel}.} On the immanent conception of empirical equivalence, hole-diffeomorphic models {are} empirically equivalent, even when $1_M$ is taken as the standard of cross-model {point} identity. To insist on this is not, \emph{pace} Weatherall, to engage in an illegitimate equivocation between $1_M$ and $\tilde{\psi}$.

When models are compared via $1_M$, Weatherall implicitly rejects the immanent conception in favour of the following criterion:
\begin{quote}
Two models are empirically distinct just in case they assign different field values to the same manifold points.
\end{quote}
(We are here assuming that there is no `internal' gauge redundancy in how field values represent physical quantities.) Given this assumption, if $g_{ab}\left(p\right) \neq \tilde{g}_{ab}\left(p\right)$, then $\mathcal{M}$ and $\widetilde{\mathcal{M}}$ are empirically distinct.

One way in which one might seek to justify this criterion is via (what we will call) the `transcendental' conception of empirical (in)equivalence, according to which the `observer' is explicitly recognized to be omitted from the models. This conception is arguably implicit in Weatherall's discussion of what an observer at $p$ would measure according to the different models.\footnote{As we have effectively already noted, this conception is not compatible with the completeness of the worlds featuring in Maudlin's discussion. It is therefore not available to Maudlin as a defence of his claims concerning observational (in)equivalence.}

Are hole-diffeomorphic models compared using $1_M$ to be regarded as not empirically equivalent according to a transcendental conception of empirical equivalence? Does this conception licence letting spacetime points or regions go proxy for the fixed perspective of a potential observer across a class of hole-diffeomorphic models? It is not clear to us that it does.

According to the interpretation of models that underwrites the transcendental conception of empirical equivalence, a model represents only a proper subsystem of the universe; the observer, in particular, is omitted from the model. It has recently been recognized that this way of thinking about models promises to provide a systematic way to adjudicate when symmetry-related models are empirically equivalent or, more generally, cannot be interpreted as representing distinct physical situations.\footnote{Our thinking here has in large part been influenced by \citetbjps{Wallace}, whose position we take ourselves to be endorsing.} For example, returning to the static shift case, it is notoriously controversial whether a substantivalist can always regard statically-shifted models of a Newtonian theory as physically equivalent. But all hands should agree that the models can be thought of as empirically distinct when taken to model a subsystem of the universe that can stand in different spatial relations to an external observer, who can detect those spatial differences using physical means (such as light) that fall outside of the physics represented in the models.
A similar story can underwrite the physical inequivalence of certain models of general relativity related by diffeomorphisms that satisfy certain non-trivial boundary conditions.\footnote{That such models should be regarded as physically inequivalent has been urged, in particular, by \citetbjps{Elvis}, albeit not for reasons to do with empirical (in)equivalence, but rather for reasons to do with the construction of conserved quantities in general relativity. Note that Belot (in contrast to Wallace) does not believe that recognizing the models' physical inequivalence requires that they be regarded as modelling proper subsystems of the universe.}

Does thinking of models related by a hole diffeomorphism as representing only subsystems of the universe allow one to regard them as empirically inequivalent? We do not see how it can. Following \citetbjps{Wallace}, the question to ask is how the symmetry relating the models extends to a set of more inclusive models that include the observer and their detection of the relevant differences. Recall that, according to Weatherall's account of the models' empirical distinctness, the unmodelled observer is imagined to be located at a spacetime $p$ on which the diffeomorphism acts non-trivially. If we now imagine including fields representing the observer and their ability to detect differences in the values of the originally modelled fields, 
it seems there are two possibilities: either the diffeomorphisms are symmetries only when acting on all the fields in the expanded model, including those representing the observer and their measurements, or the diffeomorphisms are `subsystem-local' in Wallace's terminology, and remain symmetries even when acting only on the originally modelled fields.

In the latter case, one might try to argue that the original models should be treated as {physically} inequivalent (after all, the expanded models differ over the relative location of physical fields, for example, those representing spacetime curvature and those representing an observer and their material measuring devices). However, precisely because the fields representing an observer are left untransformed when the diffeomorphism acts on the fields of the original model, the differences are unobservable (so the models remain observationally equivalent) and, moreover, any theory governing the expanded models will, very obviously, be viciously indeterministic.

It is far more plausible, therefore, to suppose that any theory governing the extended models that represent the observer will be invariant under diffeomorphisms only if they act on all the fields. But now we no longer have any basis to follow Weatherall in interpreting the map $1_M$ as identifying the fixed perspectives of potential observers across models, for in the {extended} models such observers are dragged along together with the fields that they are observing by any diffeomorphism that is a symmetry of the theory.

To sum up the conclusions of this section: Weatherall's claim that when models are compared using $1_M$ they count as empirically distinct, and that therefore the hole argument is blocked, fails because: (1) the claim leaves entirely untouched the indeterminism version of the hole argument (which does not presuppose the models are empirically equivalent, only that they are equivalent in regions outside of the hole); and (2) the claim that models count as empirically distinct is anyway suspect, even when assuming a transcendental conception of empirical equivalence.

By considering the indeterminism and underdetermination versions in turn, we take ourselves to have shown that neither version of the hole argument involves an illegitimate equivocation between $1_M$ and $\tilde{\psi}$. For anyone who remains unconvinced, however, we close this section by offering the following simple analogy.

Suppose that two identical twins, Alice and Barbara, cannot be told apart by visual inspection. Now consider the following two situations, both involving the twins standing in front of you in plain sight. In the first, Alice stands to the left of Barbara and wears a red hat, while Barbara wears a blue hat. In the second, Barbara stands to the left of Alice and wears the red hat, while Alice wears the blue hat. The twins contrive to look otherwise identical in the two situations.

We think that the following two claims about this setup are obvious and uncontroversial. The two situations are observationally indistinguishable, at least by sight: one cannot tell, as the onlooker, which of the two situations one is confronted with. Nonetheless, there is a (physical) difference between the two situations: in one it is Alice in the red hat; in the other it is Barbara.

Now consider the simple analogues of Weatherall's $1_M$ and $\tilde{\psi}$. $1_M$ corresponds to a map between the two situations that maps Alice to Alice and Barbara to Barbara. It does not preserve all qualitative (visually inspectable) features of the situation. In particular it maps someone wearing a red hat to someone wearing a blue hat. $\tilde{\psi}$ corresponds to a map between the two situations that maps Alice to Barbara and Barbara to Alice. It preserves all qualitative features but does not map each twin, as she is in one situation, to herself, as she is in the other.

According to Weatherall, the two claims that we asserted above to be obvious and uncontroversial can only be jointly accepted if one engages in an illegitimate equivocation between the two maps. If one says that the situations are observationally indistinguishable, one must be comparing them via the analogue of $\tilde{\psi}$ and so cannot, in the same breath, also say that they differ because different twins wear the two hats. If one focuses on the fact that different twins wear different hats, one is comparing them via the analogue of $1_M$, in which case one cannot at the same time recognize that the situations are observationally indistinguishable.

We hope that the unreasonableness of this position is clear without our needing to labour the point further. It may be helpful to note that the observational indistinguishability of the two situations is not a property that holds of the pair of them only {relative to}, or when compared via, (the analogue of) $\tilde{\psi}$. Rather, the existence of this map, which matches up without exception qualitatively identical features, underwrites an {absolute} property possessed by the pair of situations---a feature that holds {full stop}. This is why there is no equivocation involved in jointly attending to (i) the fact that $\tilde{\psi}$ preserves all qualitative features of the spacetime, because it is an isomorphism, but also (ii) the fact that different points are assigned different properties, precisely because $1_M$ is not an isomorphism.

\section{The Argument from Mathematical Structuralism}\label{s2}

Up to this point, we have assumed that comparing the models $\mathcal{M}$ and $\widetilde{\mathcal{M}}$ via $1_{M}$ is legitimate. As the previous section has made clear, this is the comparison presupposed in standard presentations of the hole argument, where $1_M$ is assumed, tacitly or otherwise, to track which manifold points represent the same spacetime points across models. Claims of observational indistinguishability are either irrelevant to the hole argument (when the issue is indeterminism) or do not require a problematic switch from comparison via $1_M$ to comparison via $\tilde{\psi}$. We turn now to the question of whether it is indeed legitimate to compare models via $1_M$.

Weatherall's considered view seems to be that such comparisons are not legitimate. Towards the end of his article, he canvasses three possible views that Earman and Norton might be taken to attribute to the spacetime substantivalist concerning how mathematical models represent spacetime. According to the first, the substantivalist takes spacetime to be represented by just the differentiable manifold of the models, not by the manifold plus metric. According to the second, the substantivalist takes spacetime to be represented by a structure richer than a Lorentzian manifold. The third view, which is the one that Weatherall ultimately attributes to Earman and Norton's substantivalist, involves representing spacetime by
\begin{quote}
a Lorentzian manifold (and no more), such that the manifold is understood to represent spacetime points in some sense prior to, or independently of, the value of the metric at those points [\ldots \!] The pushforward of the metric along an automorphism of the manifold would represent different property assignments to the same spacetime points, corresponding to \emph{prima facie} different physical situations. \citepbjps[p.~345]{Weatherall}
\end{quote}
Of this third option, he writes:
\begin{quote}
this view, I take it, is incompatible with the background views on mathematics described in Section~1. The problem is essentially the same as for the hole argument itself: 
this view depends on taking the identity map to provide a prior notion of when the points of two Lorentzian manifolds are the same. \citepbjps[p.~345]{Weatherall}
\end{quote}

More on the background views on mathematics that Weatherall describes in the first section of his paper in a moment. First, a brief comment on whether Weatherall is correct to attribute the third view to Earman and Norton's substantivalist.

Notoriously, Earman and Norton argued that the metric field should count as a field contained {within} spacetime, not as (part of) the characterisation of the spacetime container \citepbjps[pp.~518--19]{EN}. This suggests that the first of the views that Weatherall surveys better matches Earman and Norton's intentions. Weatherall's reasons for rejecting this possibility is that he believes it would commit Earman and Norton's substantivalist to `the radical view that any two spacetimes are equivalent as long as their associated manifolds are diffeomorphic---whether the two agree on other structure' \citepbjps[p.~344]{Weatherall}. But the view, of course, has no such consequence.  Earman and Norton's substantivalist will regard non-isometric models as representing obviously physically distinct {possibilities}, but possibilities that differ in terms of their spacetimes' having (qualitatively) different {content}. This substantivalist will nonetheless regard these possibilities as involving spacetimes that are, in terms of their intrinsic characters, qualitatively identical.\footnote{Weatherall's use of `spacetimes' to refer to mathematical models is perhaps partly responsible for generating the appearance of a problem. His allowing the stress-energy of matter to be implicitly defined by $g_{ab}$, via the Einstein field equations, also serves to blur the distinction between models differing over their characterisation of spacetime \emph{per se} and models that disagree only over its content.}

In fact, in terms of Weatherall's three-way classification, Earman and Norton substantivalist's position is best understood as a combination of views (1) and (3): one represents spacetime itself as a differentiable manifold (`and no more') but one also takes the manifold to represent spacetime points `in some sense prior to, or independently of,' the value of the metric (and that of any other field) at those points. 

Earman and Norton are in the minority in classifying the metric as content rather than as an aspect of the spacetime container. A number of authors have argued against this component of their position.\footnote{Critics of this aspect of Earman and Norton's position include Maudlin \citeyearparbjps[pp.~545--49]{Maudlin90}, Hoefer \citeyearparbjps[pp.~11--13]{Hoefer} and  Pooley \citeyearparbjps[pp.~99--101]{Pooley2006}. Part of Earman and Norton's argument rests on the claim that the metric carries (gravitational) energy and momentum. For recent discussion of this vexed topic, see for example (\citealtbjps{DuerrFB}; \citealtbjps{DuerrAR}; \citealtbjps{Read2020}).} The third of Weatherall's options is, therefore, a more faithful characterisation of an interpretative stance implicit or (often) explicit in much of the philosophical literature that followed Earman and Norton's paper. What, according to Weatherall, is wrong with it?

Near the beginning of his article, Weatherall sets out the following two interpretative principles:
\begin{quote}
        (1) our interpretations of our physical theories should be guided by the formalism of those theories; and (2) insofar as they are so guided, we need to be sure that we are using the formalism correctly, consistently, and according to our best understanding of the mathematics. \citepbjps[p.~330]{Weatherall}
\end{quote}
The application that Weatherall seeks to make of these principles focuses only on matters of sameness and equivalence. Principle (2) enjoins us to attend to our best understanding of the mathematics involved in a theory's formalism. In this connection, Weatherall endorses a view that he takes to be `a variety of mathematical structuralism' \citepbjps[fn.~8]{Weatherall}: that isomorphism is the relevant standard of sameness in mathematics.\footnote{Varieties of mathematical structuralism can differ over whether isomorphic models are understood to be (a) numerically distinct but in some sense treated `as if' identical, versus (b) numerically identical. Weatherall clearly does not assume anything as strong as (b) (cf.~\citealtbjps[p.~331, fn.~8]{Weatherall}). The `univalence axiom' of homotopy type theory affords a natural means of formalizing mathematical structuralism---see \citepbjps[p.~104]{Dougherty}. Our concerns here  also speak against any appeal to homotopy type theory as a means of evading the hole argument. For more on homotopy type theory and the hole argument, see \citepbjps{Dougherty2} and \citepbjps{LP}.}  Applying this to the case at hand, if models of general relativity are taken to be Lorentzian manifolds, isometry is the standard of isomorphism. So, hole-diffeomorphic models such as $\mathcal{M}$ and $\widetilde{\mathcal{M}}$, which  are isometric, count as `mathematically equivalent'.

Turning to principle~(1), what does the mathematical equivalence of such models imply about their physical interpretation? Weatherall's official answer is that isomorphic models should be taken to have the same `representational capacities [\ldots] if a particular mathematical model may be used to represent a given physical situation, then any isomorphic model may be used to represent that situation equally well' \citepbjps[p.~332]{Weatherall}.  We take this thesis to correspond precisely to the position that Fletcher---a self-declared ally of Weatherall on these matters---labels REME (`Representational Equivalence by Mathematical Equivalence' \citepbjps[p.~233]{Fletcher}).

We concur with Weatherall and Fletcher that REME is a legitimate principle of mathematical representation. REME, however, very obviously fails, by itself, to block the hole argument, or to rule out the allegedly problematic use of isomorphic models deployed in standard presentations of the argument. All that this use requires is that {if} model $\mathcal{M}$ (say) is taken to represent a particular possible world $W$, then $\mathcal{\widetilde{M}}$ represents (or, better, {may be taken to represent}, perhaps by further specifying the representational context) a distinct (but merely haecceitistically distinct) possibility $\widetilde{W}$. This is all completely compatible with $\mathcal{M}$ and $\mathcal{\widetilde{M}}$'s having identical representational capacities. That merely requires that $\mathcal{\widetilde{M}}$ might equally well have been chosen initially to represent $W$ (in which case, some other hole-diffeomorphic model, related to $\mathcal{\widetilde{M}}$ just as $\mathcal{\widetilde{M}}$ is related to $\mathcal{M}$, could be taken to represent $\widetilde{W}$).\footnote{Roberts \citeyearparbjps[p.~258]{Roberts} and Gryb and Th\'{e}bault \citeyearparbjps[pp~566--67]{TG} make essentially the same point. There are also parallels here with the `semantic relationalism' of Fine \citeyearparbjps{Fine, Fine2}, according to which there is no semantic difference between variables such as $x$, $y$, as illustrated by the fact that one's choice of variable in formulae such as $x>0$ or $y>0$ is purely conventional; however, there {is} a semantic difference between pairs of variables---as illustrated by the fact that $\left( x, x\right)$ is not semantically equivalent to $\left( x , y \right)$, for $x>x$ may (after quantification) express a distinct proposition from $x>y$. 
(For some critical comments on Fine's work, see \citepbjps[pp.~13-4]{BW}.)}

REME, interpreted to the letter, therefore fails to block the hole argument. Nevertheless,  it is clear that both Weatherall and Fletcher believe that a proper understanding of the mathematical formalism of general relativity {does} rule out the use of models that the hole argument presupposes. What stronger constraint on representation are they assuming?

A tempting thought might be that mathematical structuralism goes hand-in-hand with taking isomorphism equivalence classes of models to be in one-to-one correspondence with physical possibilities. This, however, is a position that Weatherall \citeyearparbjps[p.~332]{Weatherall} explicitly rejects, and for good reason. To see the problem, it is useful to break down this new thesis into two components: (i) that all mathematically equivalent models represent the same, unique possibility, and (ii) that any two mathematically inequivalent models represent distinct possibilities.

These two claims about representation correspond closely (but not exactly\footnote{While (ii) requires that mathematically inequivalent models represent distinct possibilities, Fletcher's RDMI below requires only that they do not represent the same possibility (equally well).}) to two other theses presented---and criticized---by Fletcher \citeyearparbjps[pp.~231--33]{Fletcher}:
\begin{description}
\item{RUME:} If two models of a physical theory are mathematically equivalent, then there is a unique physical state of affairs that they represent equally well.
\item{RDMI:} If two models of a physical theory are not mathematically equivalent, then it’s not the case that there is a unique physical state of affairs that they represent equally well.\footnote{RUME abbreviates `Representational Uniqueness by Mathematical Equivalence'; RDMI stands for `Representational Distinctness by Mathematical Inequivalence.'} 
\end{description}

One of Fletcher's central illustrative examples in his case against RUME and RDMI involves schwarzschild spacetimes---spherically symmetric solutions of general relativity, differing by the so-called `schwarzschild radius' $R_S$, which can be interpreted as representing appropriately isolated, non-rotating stellar bodies (for example stars, black holes, \emph{et cetera}). Against RUME, Fletcher notes that `\emph{each} mathematical Schwarzschild spacetime can represent \emph{any} physical Schwarzschild spacetime (that is, with any Schwarzschild radius) through an appropriate choice of units' \citepbjps[p.~235]{Fletcher}. Conversely, non-isomorphic schwarzschild models can represent one and the same black hole in different contexts, via a change of units, undermining RDMI \cite[pp.~237-8]{Fletcher}.

We are happy to take Fletcher's arguments against RUME and RDMI as decisive. But we note that most of these arguments exploit the fact that representation is a pragmatic matter, always involving some level of idealisation and abstraction, as well as some more-or-less arbitrary conventions, {which can shift from one representational context to another}. Fletcher correctly shows that RDMI and RUME are implausible principles if taken to hold {across} representational contexts. But that leaves open the question of whether they are legitimate principles when relativized to a fixed representational context.

In our view, the interesting questions have always been about the potential {joint} representational uses of different mathematical models: whether two models can be taken to be `co-representational' and, if so, whether they represent, when so taken, the same or different possibilities.

Occasionally, such questions have been raised explicitly in other discussions of the hole argument. In assigning a central role to the question of when two models can be understood as `co-intended', \citetbjps[p.~423]{Ryno} is investigating precisely the constraints on when two models can be understood as co-representational. Gryb and Th\'{e}bault \citeyearparbjps[pp.~566-7]{TG} draw the  distinction between what models can represent `taken in isolation' and what they can represent `taken together'. 
And Roberts homes in on a closely related question when he asks whether two isomorphic structures can have ‘co-representational capacity’ (whether they can represent the same state of affairs `at once'---a thesis he dubs `strong Leibniz equivalence'). He contrasts this question from the REME-like question of whether they have `equal representational capacity' (`weak Leibniz equivalence'), where here their ability to represent the same state of affairs need not be `at once' (that is, it need not be relative to the same representational context) \citepbjps[p.~252]{Roberts}. 
Focusing, then, on the co-representational capacities of models, what might plausible, context-relativized variants of RUME and RDMI look like? We propose the following as natural emendations of Fletcher's theses:
\begin{description}
\item{RUME$^{*}$:} If two models of a physical theory are mathematically equivalent and if one model is chosen to represent (to some degree of accuracy) a particular physical possibility (thereby fixing the representational context), then the other model, relative to that choice (that is, relative to the same representational context), also represents that physical possibility (to the same degree of accuracy).
\item{RDMI$^{*}$:} If two models of a physical theory are not mathematically equivalent and if one model is chosen to represent (to some degree of accuracy) a particular physical possibility, then the other model, relative to that choice, does not represent that physical possibility (to the same degree of accuracy).
\end{description}
To these we add a third thesis---`Mathematical Inequivalence follows from Representational Distinctness':
\begin{description}
\item{MIRD$^{*}$:} If, relative to some representational context, two models of a physical theory represent distinct physical possibilities, then they are not mathematically equivalent.
\end{description}

Before assessing these theses, the logical relationship between MIRD$^{*}$ and RUME$^{*}$ should be noted. MIRD$^{*}$ is equivalent to the claim that if two models of a physical theory are mathematically equivalent and one of the models represents a particular physical possibility relative to some representational context, then the other model, relative to that context, {either} represents the same possibility {or fails to represent a possibility}. It therefore corresponds to a natural weakening of RUME$^{*}$, where the original consequent is replaced by a disjunction of that original together with an alternative whose salience only becomes apparent when one is considering the context-relativized principles.

One consequence of this is that someone denying RUME$^{*}$ need not, thereby, be committed to denying MIRD$^{*}$. One might deny RUME$^{*}$ and affirm MIRD$^{*}$ because one thinks that, if one of two isomorphic models is chosen to represent a particular possibility then the other, though equally apt to represent that possibility (as per REME), acquires the status of an `unintended interpretation' \citepbjps[p.~420]{Ryno}. One is again reminded of lost insights from the discussion that flourished in the immediate aftermath of Earman and Norton's paper. \citetbjps{Butterfield} lays out essentially the same three options in the following terms. First he asks of two isomorphic models: do they represent the same possibility? (This is his question `(Same?)' \citepbjps[p.~12]{Butterfield}.) If the answer is No, one option is that they represent distinct possibilities (Butterfield's `(Each)'). But one might also hold that only one model may represent a possibility relative to any given representational context (cf.~Butterfield's `(One)').

Fletcher, in motivating his discussion of RUME, RDMI and REME, seeks to attribute to a number of authors a misguided commitment to RUME. In our view, the more charitable reading of much of the literature to which Fletcher adverts sees it as committed to the (weaker) starred variant (or to a close relative thereof). In particular, Earman and Norton's `Leibniz Equivalence' states that `diffeomorphic models represent the same situation' \citepbjps[p.~522]{EN}. Surely, once the need to attend to the representational context is taken into account, this is most plausibly interpreted as RUME$^*$, not as RUME. Such a reading makes good sense of the alleged conflict with substantivalism. We have seen that the interpretation of hole-diffeomorphic models to which the substantivalist is allegedly committed to acknowledging as legitimate, that is, as {jointly} representing possibilities that differ merely haecceitistically, is not in conflict with REME. It is in conflict with RUME$^*$ because (and here we finally get the heart of the matter) it {directly} conflicts with MIRD$^*$, which, as we noted, is entailed by RUME$^*$.

The starred theses might make better sense of much of the now-canonical discussions, but do any of them plausibly follow from Weatherall's two principles concerning how mathematics should guide interpretation? RUME$^*$, or Leibniz Equivalence, corresponds to the thesis which Roberts \citeyearbjps{Roberts} dubs `strong Leibniz equivalence'.\footnote{In fact, Roberts' `strong Leibniz equivalence' states only that isomorphic models \emph{can} represent the same physical state of affairs `at once'. While weakening `represent' to `can represent' is well-motivated for a cross-context principle, favouring REME over RUME, it is less obviously motivated when one is relativizing to a representational context. This is another reason to take traditional (unmodalized) statements of Leibniz Equivalence to be implicitly context-relativized and so akin to RUME$^*$ (rather than as expressing an obviously unmotivated principle like RUME).} As such, it is vulnerable to the telling criticisms to which he subjects it. But, as we have seen, while RUME$^*$ entails MIRD$^*$, it is not equivalent to it. Arguably it is this latter thesis that deserves to be the focus of attention and, indeed, Fletcher comes close to giving an explicit argument for it. He writes:
\begin{quote}
Lorentzian manifolds may not exemplify [all the] properties of the states of affairs they represent, but all the properties they \emph{do} exemplify---those not abstracted away---are the same for isomorphic manifolds. This is precisely encoded in the mathematical models themselves with the interpretation of isomorphic objects in a mathematical category as being equivalent \emph{as objects in that category}. \citepbjps[p.~239]{Fletcher}
\end{quote}
Put slightly differently, two situations are distinct if one possesses a feature that the other lacks. So for two models jointly to represent two distinct situations (and to represent them as distinct) one model must represent one situation as possessing a feature that the other model represents the other situation as lacking. But, if we are treating the models as objects of a certain category, then this representational difference has to correspond to a difference between them considered as objects of that category. Since isomorphic models are precisely those models that do {not} differ as objects of the category, such models cannot (if really being treated as objects of that category) differ representationally in the manner required.

That, at any rate, is our best effort at reconstructing the argument from mathematical structuralism. We finish with three critical observations concerning where we think this leaves the debate.

First, it bears stressing how natural and apparently appropriate is the particular use of isomorphic models that Weatherall and Fletcher criticize. A substantivalist who believes that possible spacetimes can differ merely haecceitistically is after a way to represent the structure of such a spacetime, which (let us suppose) really is no richer than that of a Lorentzian manifold, and at the same time talk about two possibilities that exemplify the very same structure but differ over which individuals possess which particular properties. How better to do this than to use isomorphic models of the appropriate structural type that differ merely over which of the base elements of their sets are assigned the structural properties common to both models?

The second of the three options Weatherall canvassed for Earman and Norton's substantivalist was to represent spacetime as having a structure richer than that of a Lorentzian manifold. Fletcher talks of including `spacetime point haecceities' \citepbjps[p.~239]{Fletcher}. But why should the substantivalist be saddled with such baroque machinery in order to have two models that can stand for the different possibilities that they acknowledge when these possibilities differ {merely} over which individuals have which properties? Misleading terminological etymology notwithstanding, haecceitists need not be committed to haecceities.

Our second observation is that Fletcher himself, in one of his arguments against RUME, endorses exactly this type of {co-representational} use of isomorphic models to represent distinct possibilities! His example involves a collection of structures containing nothing but a Minkowski metric on a manifold diffeomorphic to $\mathbb{R}^4$, together with an inextendable, `jointed' timelike curve composed of an initial geodesic segment followed by a segment exemplifying constant proper acceleration. He holds that this collection of structures can represent a variety of possibilities for a lone particle that ceases, at some moment, to move inertially and starts to accelerate. The possibilities in question are supposed to differ merely over when and in which direction the particle starts to accelerate. After taking a particular model to represent one such a history, Fletcher writes:
\begin{quote}
Users of relativity theory, I claim, would intend for the theory to endorse that the particle could have swerved (at the same acceleration) in another direction, even at another time, than it did in the above model. Such alternative states of affairs could be realized by a spatial rotation or time translation acting on the above model. Yet the resulting model is in fact isomorphic to the one above. \citepbjps[p.~237]{Fletcher}
\end{quote}

In order for these models to (co-)represent states of affairs that differ only over when and in which direction the particle swerves, a standard of cross-model point identity needs to be assumed (as Fletcher himself spells out) which, although an isometry of the metric substructure, is not, \emph{pace} Fletcher \citeyearparbjps[p.~240]{Fletcher}, an isomorphism of the relevant category (that is, that of `Minkowski spacetimes with distinguished worldlines'). Fletcher claims that the inclusion of a distinguished worldline in the example serves to differentiate the case from that of the hole argument. We, by contrast, take its inclusion to make the cases precisely analogous. In both cases, the co-representational function of the models presupposes comparing isomorphic models via a map that is not (or need not be) an isomorphism. And in both cases the various physical states of affairs supposedly represented by different isomorphic models differ merely haecceitistically. (Lone particle worlds that differ only over when and in which direction the particle swerves are worlds that differ only over which spacetime points have the property of being a location of the particle.)

Our first two observations have questioned whether MIRD$^*$ should be accepted on the basis of mathematical structuralism. Our third and final observation disputes that embracing MIRD$^*$ {on that basis} blocks the hole argument.

One theme of Weatherall's paper is that the formalism of general relativity does not {generate} a philosophical problem. We agree with this. One is not forced to recognize a plurality of physical possibilities merely because a theory admits as solutions a plurality of isomorphic models. But equally, one does not rule out the existence of such a plurality merely by prohibiting a particular interpretation of the formalism.

For Earman and Norton, and the literature they unleashed, the problem was not the formalism of general relativity \emph{per se}, but {substantivalism}.\footnote{Which means that the formalism is responsible for the problem {indirectly}, to the extent that it leads to the postulation of spacetime points.} For those who accept Earman and Norton's `acid test' of substantivalism, possibilities involving a common pattern of geometrical properties and a common material content can differ merely over which substantival points possess which properties.

Suppose that $C$ is a class of qualitatively identical worlds---and make no assumptions (Earman and Norton's `acid test' notwithstanding) about the cardinality of $C$, which might perhaps contain only a single member. Let $\mathcal{M}$ and $\widetilde{\mathcal{M}}$ both be members of a class of isometric spacetime models of a type and particular character that makes them apt to represent members of $C$.\footnote{So, in line with REME, each model can serve equally well to represent any element of $C$, however many there are. As we have seen, this does not block the hole argument.} Someone who embraces MIRD$^*$ for Weatherall's and Fletcher's reasons disavows being able to use $\mathcal{M}$ and $\widetilde{\mathcal{M}}$ to jointly describe different members of $C$ (if different members there are). But they do not thereby save general relativity from indeterminism. If $C$ really does contain a plurality of members differing in the way envisaged by some substantivalists, then general relativity according to Weatherall and Fletcher does not distinguish between the possibilities even to the extent of not being able to refer differentially to them. It therefore  (implicitly) regards them as all equally possible, which is just to say that, for the relevant type of substantivalist, the theory is indeterministic.

All hinges on whether, on the assumption of substantivalism, $C$ does indeed involve the relevant plurality. One cannot answer that question without engaging with the themes laid bare in the orthodox philosophical literature.\footnote{Cf.~\citepbjps[pp.~388--89]{Teitel}.} One could see here a partial vindication of what \citetbjps{TMN} dubs the `motivational' approach to symmetries over the `interpretational' approach.\footnote{See also (\citealtbjps{NMJR}; \citealtbjps{JRTMN}; \citealtbjps{TMNJR}).} In accepting MIRD$^*$ for formal reasons, Weatherall and Fletcher exemplify an `interpretational' approach, on which symmetry-related models of a given theory can be ruled out \emph{ab initio} as (co-)representing distinct physical states of affairs.\footnote{Indeed, they celebrate their lack of engagement with metaphysics: see \citepbjps[pp.~330, 343--45]{Weatherall} and \citepbjps[p.~240]{Fletcher}.} In contrast, the `motivational' approach would hold that 
MIRD$^*$ should only be adopted once it has been secured by what M\o{}ller-Nielsen \citeyearparbjps[p.~1256]{TMN} calls a `metaphysically perspicuous characterisation' of the theory's ontology. Only by engaging with the metaphysics of space and time can one confront the question of whether there is or is not a plurality $C$ of worlds corresponding to an isomorphism class of models.

\section{Conclusions}

There is much in Weatherall's paper with which we agree---in particular, the claims that (i) hole-diffeomorphic models of general relativity have identical representational capacities, and (ii) that mathematical structuralism can be taken to show that the formalism of general relativity does not {generate} a philosophical problem. This notwithstanding, these observations are insufficient to block the hole argument---for the problem can be posed without equivocation on the use of $1_M$ versus $\tilde{\psi}$, and, ultimately, any serious attempt to grapple with this problem must involve a careful engagement with metaphysics.

\section*{\centering Acknowledgements}

We are very grateful to Neil Dewar, John Dougherty, Sam Fletcher, Tushar Menon, Tom M\o{}ller-Nielsen, John Norton, Bryan Roberts, and especially Jim Weatherall, for helpful discussions on this material. We also thank the anonymous referees for helpful comments. J.R.~acknowledges support from the John Templeton Foundation (grant number 61521).

\vspace{0.5cm}

\begin{flushright}
{\em {\em Oliver Pooley} \\
Faculty of Philosophy \\
University of Oxford \\
Oxford, UK \\
oliver.pooley@philosophy.ox.ac.uk
\vspace{0.5cm}
\\
{\em James Read} \\
Faculty of Philosophy \\
University of Oxford \\
Oxford, UK \\
james.read@philosophy.ox.ac.uk}
\end{flushright}

\end{document}